\documentclass{article}
\usepackage{emulateapj}
\usepackage{apjfonts}

\begin{document}

\submitted{To appear in The Astrophysical Journal Letters}

\title{Far-Ultraviolet Emission from Elliptical Galaxies at
$\lowercase{z}=0.33^1$}

\author{Thomas M. Brown, Henry C. Ferguson, Ed Smith}

\affil{Space Telescope Science Institute, 3700 San Martin Drive,
Baltimore, MD 21218.  tbrown@stsci.edu, ferguson@stsci.edu,
edsmith@stsci.edu}

\vskip 0.1in

\author{Charles W. Bowers, Randy A. Kimble}

\affil{Code 681, NASA Goddard Space Flight Center, Greenbelt, MD 20771. 
charles.w.bowers@nasa.gov, randy.a.kimble@nasa.gov} 

\author{Alvio Renzini}

\affil{European Southern Observatory, Karl-Schwarzschild-Strasse 2, 
Garching bei M$\ddot{\rm u}$nchen, Germany arenzini@eso.org} 

\author{R. Michael Rich}

\affil{Division of Astronomy, Dpt.\ of Physics \& Astronomy, UCLA, Los
Angeles, CA 90095. rmr@astro.ucla.edu}

\begin{abstract}

We present far-ultraviolet (far-UV) images of the rich galaxy cluster
ZwCl1358.1+6245, taken with the Space Telescope Imaging Spectrograph
on board the Hubble Space Telescope (HST).  When combined with
archival HST observations, our data provide a measurement of the
UV-to-optical flux ratio in 8 early-type galaxies at $z=0.33$.
Because the UV flux originates in a population of evolved, hot,
horizontal branch (HB) stars, this ratio is potentially one of the
most sensitive tracers of age in old populations -- it is expected to
fade rapidly with lookback time.  We find that the UV emission in
these galaxies, at a lookback time of 3.9 Gyr, is significantly weaker
than it is in the current epoch, yet similar to that in galaxies at a
lookback time of 5.6 Gyr.  Taken at face value, these measurements
imply different formation epochs for the massive ellipticals in these
clusters, but an alternative explanation is a ``floor'' in the UV
emission due to a dispersion in the parameters that govern HB
morphology.

\end{abstract}


\section{Introduction}

Elliptical and S0 galaxies show a striking rise in their spectra at
$\lambda < 2000$~\AA.  Spectroscopy (Brown et al.\ 1997) and imaging
(Brown et al.\ 2000b) of local galaxies show this ``UV upturn'' arises
from a minority population of hot horizontal branch (HB) stars. On
energetic grounds, these evolved stars were long considered the best
candidates for the UV emission (Greggio \& Renzini 1990), with the
implication that the UV-to-optical color could be the most rapidly
evolving feature in the spectra of ellipticals.  In theory, the UV
upturn can fade by several magnitudes as the lookback time increases
by a few gigayears, but the evolution is very model-dependent, and a
strong UV upturn could appear at ages as early as $\sim 6$ Gyr
(Tantalo et al. 1996) or as late as $\gtrsim$15 Gyr (Yi, Demarque, \&
Oemler 1998).

This evolution has motivated our ongoing survey to measure the UV
upturn in clusters at intermediate redshifts.
Previously, we observed the
rich clusters Abell 370 at $z=0.375$ (Brown et al.\ 1998) and
CL0016+16 at $z=0.55$ (Brown et al.\ 2000a); those observations
implied that there was little fading of the UV upturn out to $z \approx 0.4$
with a significant decline at higher $z$.
However, further cluster measurements are clearly needed to
explore the universality of this evolution.

To that end, we have obtained far-UV images of ZwCl1358.1+6245 with
the Space Telescope Imaging Spectrograph (STIS) on board the Hubble
Space Telescope (HST).  This is a rich, compact cluster of galaxies
originally identified by Zwicky \& Herwig (1968) and later
rediscovered in the x-\linebreak

{\small \noindent $^1$Based on observations made with the NASA/ESA Hubble
Space Telescope, obtained at the Space Telescope Science Institute,
which is operated by AURA, Inc., under NASA contract NAS
5-26555. These observations are associated with proposal 8564.}

\noindent
ray observations of Luppino et al.\ (1991).  It
lies at intermediate redshift ($z=0.33$; Fisher et al.\ 1998)
with little foreground extinction ($E_{B-V} = 0.023$ mag;
Schlegel, Finkbeiner, \& Davis 1998).  When combined with archival
images from the Wide Field Planetary Camera 2 (WFPC2) on HST, our
far-UV images provide a measurement of the UV upturn for 8 elliptical
and S0 galaxies in the cluster core. Seven of these galaxies are
well-detected ($> 3 \sigma $) in the far-UV, and the central galaxy
shows strong extended UV emission, likely associated with infalling
matter.  The weak UV upturn measured in these galaxies is in contrast
to the strong UV upturn measured previously at $z=0.375$, yet
surprisingly similar to that at $z=0.55$.

\section{Observations and Data Reduction}

\subsection{Far-UV Imaging}

We imaged ZwCl1358.1+6245 for 68,291 s while the cluster was in the
HST Continuous Viewing Zone.  The field was centered at an R.A.\ of
$13^h59^m50.63^s$ and a Dec.\ of $62^{\rm o}31\arcmin 5.3\arcsec$
(J2000) with a position angle of 50$^{\rm o}$, placing 8 S0 and
elliptical galaxies within the STIS $25\arcsec \times 25\arcsec$ field,
including the brightest galaxy of the cluster (see
Figure 1). Exposures were dithered by 1--2 pix to allow removal of hot
pixels.  We used the F25QTZ filter (1450--1900~\AA) to
minimize the background from geocoronal emission in \ion{H}{1}
$\lambda 1216$ and \ion{O}{1} $\lambda 1304$.  Note that the
sensitivity of the far-UV camera has been slowly decreasing
at the rate of $\sim$1.5\% yr$^{-1}$; we include this decline in our
analysis (at the time of our observations, a flat
spectrum of $1.19 \times 10^{-16}$ erg$^{-1}$ s$^{-1}$ cm$^{-2}$
\AA$^{-1}$ produced 1 count s$^{-1}$ in this bandpass).

The STIS far-UV detector has a very low dark rate when cold ($\sim 6
\times 10^{-6}$ cts sec$^{-1}$ pix$^{-1}$), but as observations
progress, a dark count ``glow'' appears, centered in the upper
left-hand quadrant of the detector.  Using the optical images to mask
objects, we normalized and subtracted from each exposure a profile of
this dark glow (created from a sum of $\sim 500$~ks of dark
exposures).  We co-added the exposures using the IRAF DRIZZLE package,
weighting the pixels in each frame by the ratio of the exposure time
squared to the dark count variance, including a hot pixel mask.  The
algorithm weights the exposures by the square of the signal-to-noise
ratio for sources fainter than the background.  Note that the STIS UV
detectors are photon counters that register less than 1 count per
cosmic-ray hit, and thus the images do not require the cosmic-ray
rejection required for processing CCD images.

\vskip 0.1in \hskip 0.5in
\parbox{2.5in}{\epsfxsize=2.5in \epsfbox{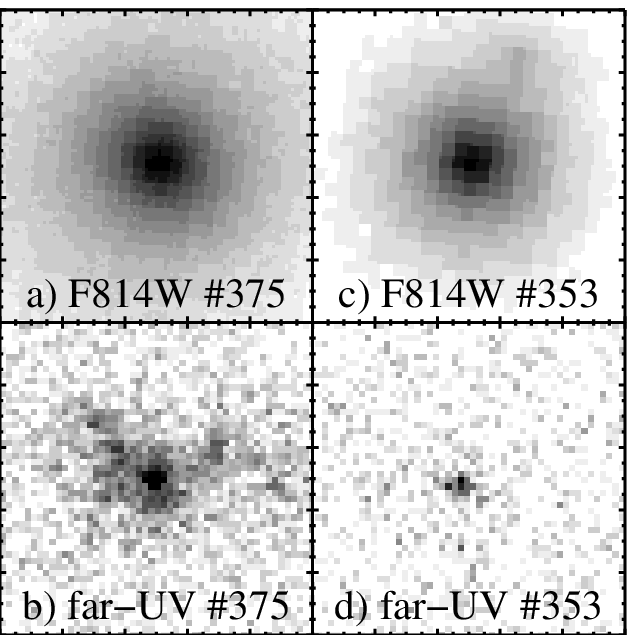}}

\vskip 0.1in
\parbox{3.25in}{\small {\sc Fig.~1--}
Optical and far-UV
images of two early-type galaxies in ZwCl1358.1+6245 ($2.5\arcsec
\times 2.5\arcsec$).  Ellipticals at low $z$
usually appear compact in the UV, but the brightest cluster member
(panels $a$ \& $b$) shows irregular structure that is not seen in the
optical, likely from Lyman-$\alpha$. The other galaxies in our sample
appear compact in the UV, as in panel $d$.}

\subsection{Optical Data}

We used the DRIZZLE package to co-add the archival WFPC2 images of this
cluster, reject cosmic rays and hot pixels, and register the co-added
optical data to the far-UV image.  WFPC2 data in the F606W (broad $V$)
and F814W ($I$) bandpasses were available for our entire field; these
data come from an impressive mosaic of the cluster by van Dokkum et
al.\ (1998), who also provide morphological classifications for the
galaxies in our sample (Table 1).  Partial coverage of our field was
also available in F450W ($B$).  The UV upturn strength is usually
measured by the rest-frame $m_{1550}-V$ color, where the flux from
1250--1850~\AA\ gives $m_{1550}$ and $V$ is Johnson $V$.  Our far-UV
bandpass corresponds to rest-frame 1090--1429~\AA\, while the F814W
bandpass corresponds roughly to Johnson V in the rest-frame.  In \S 2.5,
we will use the spectral energy distributions of local
elliptical galaxies, redshifted to $z=0.33$, to interpret the observed
count rates in these bandpasses as $m_{1550}-V$ colors.

\subsection{Extended Emission}

Due to their low luminosity and sharp profile in the far-UV,
elliptical galaxies usually appear point-like in images of distant
clusters.  What is striking about Figure 1 is the appearance of the
brightest cluster member, which shows extended tenuous emission that
is not seen in the broad WFPC2 bandpasses at longer wavelengths.
The cluster shows significant x-ray luminosity (Stocke et
al.\ 1991), and associated H-$\alpha$ emission (Donahue, Stocke, \&
Giola 1992).  The H-$\alpha$ contours extend outward from the nucleus
in three prongs, and two of these trace the far-UV emission when the
Donahue et al.\ (1992) image is aligned to our own, strongly
suggesting that the far-UV emission is due to Lyman-$\alpha$
redshifted into the STIS far-UV bandpass.  The x-ray and H-$\alpha$
emission were historically taken as evidence of a ``cooling flow,''
although the presence of cooling gas is now controversial, and this
term may be a misnomer.  Lyman-$\alpha$ emission is expected to trace
the H-$\alpha$ emission in galaxy clusters, whether or not the
infalling material is cooling in the historical sense, but it is
rarely seen because this emission is lost in the geocoronal
Lyman-$\alpha$ emission when observing local clusters.  Unfortunately,
both the Lyman-$\alpha$ emission and the stellar light peak toward the
center of this galaxy, so it is difficult to separately quantify the
two.

Another, less likely, source would be star-formation from the
infalling material.  Normalizing a model spectrum of constant
star-formation to the far-UV emission, we find that such star
formation would be difficult to detect in the optical
images. In any case, the optical images show no evidence of the
morphology seen in the far-UV. \\

\noindent
\parbox{3.0in}{
{\sc Table 1:} Photometry

\begin{tabular}{ccccc}
\tableline
         &   &  far-UV             & F814W    & $m_{1550}-V$ \\
ID$^{a,b}$& Morph.$^a$ &(cts)              & (DN)     & (mag)    \\
\tableline
381 & E  & $ 147 \pm 27$      &$33868\pm 71$&$4.6\pm 0.2$\\
375 & E  & $ 2873 \pm 66$     &$52928\pm 89$&$1.9^c\pm 0.1$ \\
368 & S0  & $ 163 \pm 49 $     &$13965\pm 50$&$3.5^{+0.4}_{-0.3}$ \\
357 & E  & $ -23 \pm 52$      &$22068\pm 57$&$> 5.9$ \\
360 & E  & $ 139 \pm 47$      &$20255\pm 55$&$4.1\pm0.4$ \\
358 & S0  & $ 149 \pm 41$      &$17587\pm 51$&$3.8^{+0.4}_{-0.2}$ \\
354 & E  & $ 118 \pm 33$      &$17165\pm 51$&$4.1\pm 0.3$ \\
353 & E  & $ 405 \pm 29$      &$48531\pm 92$&$3.9 \pm 0.1$ \\

\tableline
\end{tabular}
$^a$ Fisher et al.\ (1998).\\
$^b$ van Dokkum et al.\ (1998).\\
$^c$ Contaminated by Lyman-$\alpha$ emission.\\
}

\subsection{Photometry}

We performed aperture photometry on the far-UV image using our own IDL
software, with a source aperture of radius 16 pixels (0.4$\arcsec$)
and a sky annulus of radii 80 and 100 pixels, as done previously by
Brown et al.\ (2000a).  Each aperture was centered on the bright core
of the galaxy, determined from the F814W WFPC2 data.  Statistical
errors for the photometry include the Poisson contribution from the
source counts, the statistical uncertainties in the spatially-varying
background, and the effects of the weighted co-addition of the frames
(see the previous section).  Using the DAOPHOT package in IRAF, we
measured the flux in the F814W data (registered via DRIZZLE to the
far-UV frame) using the same source and sky apertures.  The optical
and far-UV photometry are shown in Table 1.

Note that the aperture size does not contribute significant systematic
errors in our analysis, whether we are considering measurements at
different $z$ or in different bandpasses.  Except for galaxy
375 (which has extended emission), there is little variation in our
$m_{1550}-V$ colors as we reduce the aperture size -- they remain
uniformly red, with small variations consistent with the statistical
uncertainties in Table 1.  Measurements of $m_{1550}-V$ in local
ellipticals (Burstein et al.\ 1988) were performed with a
$10\arcsec \times 20\arcsec$ aperture, while our 0.8$\arcsec$ diameter
aperture would subtend $\sim 50\arcsec$ at the distance of Virgo
(characteristic of the Burstein et al.\ sample).  However, in local
galaxies, there is almost no variation in the surface $m_{1550}-V$
color out to a radius of $\sim 20\arcsec$ (Ohl et al.\ 1998), and the
colors within any aperture are dominated by the core (given the
sharply peaked profiles).  We also note that in the HST
bandpasses, our aperture gives encircled energy agreement at the 5\%
level for point sources and better agreement for extended sources
(Robinson 1997$^2$; Holtzman et al.\ 1995).

\subsection{UV-to-Optical Colors}

The UV upturn is traditionally characterized by the rest-frame
$m_{1550}-V$ color (see Burstein et al.\ 1988).  We used the spectra of
three local elliptical galaxies (NGC1399, M60, and M49) to convert the
observed STIS and WFPC2 countrates to rest-frame $m_{1550}-V$, as done
previously in Brown et al.\ (1998) and Brown et al.\ (2000a).  Our
spectra of NGC1399, M60, and M49 have respective $m_{1550}-V$ colors
of 2.05, 2.24, and 3.42 mag (Burstein et al.\ 1988); despite the substantial 
range in UV upturn strength, the spectral shape within the far-UV range or 
within the optical range varies little from  galaxy to galaxy. We
redshifted these local templates to $z=0.33$ and then applied a
foreground reddening of $E(B-V) = 0.023$ mag (Schlegel et al.\ 1998).
We then used the IRAF CALCPHOT routine to calculate the relative
countrates for these templates in the STIS/far-UV and WFPC2/F814W
bandpasses.  The ratios of far-UV to F814W countrates observed in the
ZwCl1358.1+6245 galaxies ($R_{obs}$) were then compared to the ratios
from the templates ($R_{template}$), and the template with the closest
ratio was used to determine the rest-frame $m_{1550}-V$ (Table 1),
using the relation\\ \\ $(m_{1550}-V)_{obs}~=~(m_{1550}-V)_{template}
- 2.5{\rm log}(R_{obs}/R_{template})$.\\ \\
\noindent
Except for galaxy 375, all of the galaxies show weak UV upturn
emission.  Galaxy 375 is contaminated by the Lyman-$\alpha$ emission
discussed previously, so the UV upturn emission is really an upper
limit on the emission from the evolved HB population.  Note that in a
much smaller aperture (radius 5 pixels) that excludes much of the
Lyman-$\alpha$ emission, this galaxy has an even bluer $m_{1550}-V$ of
1.6 mag.

We show in Figure 2 the
$m_{1550}-V$ colors measured in galaxy clusters to date.  Nearby quiescent
ellipticals have been measured by Burstein et al.\ (1988) in Virgo,
Coma, and Fornax.  The galaxies at $z=0.33$ are from the present work
(excluding galaxy 375).  Measurements at $z=0.375$ are from Faint
Object Camera (FOC) observations of Abell 370 (Brown et al.\ 1998).
Measurements at $z=0.55$ are from STIS observations of CL0016+16
(Brown et al.\ 2000a).  At each redshift, the observed fluxes have
been transformed to rest-frame $m_{1550}-V$ using the spectra of local
elliptical galaxies.  

Before proceeding, it is worth noting that the $z=0.375$ data
may have significant systematic errors -- the calibration on the
FOC was far less certain than it is with STIS.  The measurements at $z
\approx 0$, $z=0.33$, and $z=0.55$ were all done with distinct
photometry in the far-UV and optical, whereas the FOC measurements
come from the ratio of two long-pass filters (see
Brown et al.\ 1998).  It would be prudent to verify the $m_{1550}-V$
colors at $z=0.375$ with a true solar-blind instrument, such as STIS
or the far-UV channel on the Advanced Camera for Surveys, to 
verify that the UV upturn is so strong in Abell 370.


\vskip 0.1in

{\small $^2$R.\ Robinson 1997, Examining the STIS Point-Spread 
Function, http://hires.gsfc.nasa.gov/stis/postcal/quick\_reports.html}

\section{Discussion}

To place our observations in context, we have plotted in Figure 2 the
models of Tantalo et al.\ (1996), which were constructed assuming gas
infall for the chemical evolution, and under the assumption that
massive ellipticals form by monolithic collapse.  The onset of the UV
upturn is very sensitive to the formation redshift ($z_f$) assumed,
and for the moment, we will consider this at face value.

Leaving aside the uncertain result at $z=0.375$, Figure 2 demonstrates
that the UV upturn does indeed fade with increasing $z$, but does so
much more slowly than, for example, in the models of Tantalo et al.\
(1996). In fact, very little $m_{1550}-V$ evolution appears to exist
between $z=0.33$ and 0.55.  Given the $m_{1550-V}$ range observed
at $z=0$, correlated with metallicity, the trend is most
apparent if one considers both the mean and the most blue $m_{1550}-V$ 
in each epoch; there are enough galaxies to
imply that the UV upturn is systematically weaker at $z=0.33$ and
0.55.  The figure suggests that galaxies in the $z=0.33$ cluster
formed at $z\sim 2$, while those in the $z=0.375$ and 0.55 clusters
formed at $z\sim 4$.  Given the evidence for hierarchical merging,
such variation may not be inconceivable.  However, we have targeted
the massive ellipticals in the centers of rich clusters, with no
spectroscopic evidence of recent star formation.  The UV upturn traces
the age of the oldest stars in these populations, not necessarily the
ages of the galaxies.  If these galaxies formed via monolithic
collapse, the age of the oldest stars reflects the age of the galaxies
themselves, but if the galaxies formed via hierarchical merging, the
oldest stars may predate the merger events and still comprise the bulk
of the population.  Studies of clusters out to $z \sim 1$ indicate
that most of the star formation in ellipticals had to be
completed at $z \gtrsim 3$, followed by quiescent
evolution (Stanford, Eisenhardt, \& Dickinson 1998; Kodama
et al.\ 1998).  Therefore, we expect little scatter in $z_f$.  In
contrast, the strong distinction between the measurements at $z=0.33$
and 0.375, and, more importantly, the similarity between the
measurements at $z=0.33$ and 0.55, implies a surprisingly large
dispersion in the formation epochs between clusters ($z_f \approx
2-4$).  Such variation in $z_f$ would be more plausible if the
nonmonotonic evolution in $m_{1550}-V$ implied by the FOC measurements
were independently confirmed.

Another (less likely) explanation for the variation in Figure 2 is
that there is a small dispersion in $z_f$ among the clusters, but the
chemical evolution varied significantly from cluster to cluster.  The
onset of the UV upturn in the Tantalo et al.\ (1996) models depends
upon the details of the chemical evolution, and the UV upturn in local
ellipticals correlates strongly with metallicity, and to a lesser
extent, with luminosity (Burstein et al.\ 1988).  However, all of the
galaxies in these programs are massive ellipticals in the centers of
rich, massive clusters -- it seems unlikely that their populations
would be chemically distinct from their analogs in the local Universe.
Even so, it would be interesting to obtain metallicity measurements
for our sample, to look for trends like those seen locally.

Models for the evolution of elliptical galaxies are sensitive not only
to $z_f$, but to other parameters, such as time of onset for galactic
winds, accretion timescale, and efficiency of star formation.  Tantalo
et al.\ (1996) tuned these parameters to reproduce the properties of
low-$z$ ellipticals, and there is not much freedom to tune them
further to match our observed UV evolution.  It is therefore worth
looking at the parameters that affect HB morphology in particular. Hot
HB stars have lost nearly all (but not all) of their envelope due to
mass loss on the RGB.  Tantalo et al.\ (1996) make specific
assumptions about the relation between mass loss and chemical
abundance in order to approximate the UV upturn range at $z=0$. 
In their models, hot HB stars do not exist at higher $z$ because their main
sequence progenitors have larger masses (resulting in redder HB
stars).  What if, instead, we imagine that at $z=0$ there is a
reservoir of stars in ellipticals that have lost so much mass
that they fail to ignite He?  See, e.g., Figure 9 in Greggio \&
Renzini (1990).  These stars could, for example, come from the
high-metallicity tail of the metallicity distribution.  At higher
$z$, the more massive stars in that metallicity range would
become hot HB stars, providing the UV flux that would otherwise be
missing.  Therefore, moving from zero to high redshift, the UV upturn
would be produced by stars at progressively higher metallicity, until
it disappears entirely when the very end of the metallicity
distribution is reached.  In more general terms, the same qualitative
behavior is expected if the occurrence of hot HB stars is driven by a
distribution of stellar ages, or a distribution of RGB mass loss, or a
combination thereof.  It is plausible that the details of the
metallicity distribution, and the relation between mass loss and RGB
parameters, could be tuned to reproduce the gradual $m_{1550}-V$ evolution
observed, with a common $z_f \sim 4$, without violating other constraints on 
stellar evolution.

Further alternatives are also possible. At $z=0$, hot HB stars may be
the dominant, but not the sole, contributor to the UV upturn.  While
the hot HB would rapidly disappear with redshift, the $m_{1550}-V$
color would initially drop, but then hit a {\it floor} value as the
role of dominant UV contributor is assumed by another class of
UV-bright objects whose evolution with redshift is not as fast as that
of the hot HB stars.  Possibilities include a low level of ongoing
(massive) star formation, and various kinds of binaries such as, e.g.,
post-RGB binary components and accreting white dwarfs (Greggio \&
Renzini 1990).  
If future observations can show that HB stars are not the sole 
contributor to the UV upturn at $0.3 \lesssim z \lesssim 0.6$, the Tantalo 
et al.\ (1996) models would imply $z_f \lesssim 2$; however, given that 
Kodama et al.\ (1998) and Stanford et al.\ (1998) find the bulk of the 
stellar population formed at $z_f \gtrsim 3$, such a finding would more 
likely tell us that the onset of the UV upturn takes place at an older
age than the $\sim6$~Gyr given by the models of Tantalo et al.\ (1996).

It would be difficult to rule out a low star formation rate (SFR) as
the source of the $m_{1550}-V$ floor.  Assuming constant star formation
with a flat $f_\nu$ spectrum, the Kennicutt (1998) relation gives 
SFR=1.4$\times 10^{-28} L_\nu \approx$ 0.005--0.02 $M_\odot$ yr$^{-1}$ for 
our $z=0.33$ sample
(excluding galaxies 375 and 357).  This is a lower limit, given that
the real spectrum would likely have a far-UV downturn due to
extinction.  The rates in the galaxies at $z=0.55$ are similar
(0.008--0.015 $M_\odot$ yr$^{-1}$).  Thus, from $z=0.6$ to $z=0.3$ (a
span of 2.4 Gyr), this SFR would produce $\sim 3\times 10^7$ $M_\odot$
of stars, which would be at least 3.6 Gyr old today if the star
formation stopped at $z=0.3$.  A remnant intermediate-age population
comprising a few percent of the total stellar mass in a normal
elliptical galaxy would be difficult to detect.  The same arguments
apply if the star formation is episodic, because the number of $> 3
\sigma$ detections at $z=0.33$ (excluding galaxy 375) and $z=0.55$
implies a duty cycle of $\sim$80\%.  This hypothesis could be tested
by searching for low levels of H-$\alpha$ or \ion{O}{2} emission.

In summary, our findings are in broad agreement with the expectation
of a fading UV upturn with redshift, due to a progressively smaller
number of hot HB stars being produced in younger populations.
However, the rate of this fading remains to be understood -- whether
the whole trend can be attributed to the redshift evolution of the
number of hot HB stars, or whether an additional class of hot stars is
contributing a floor to the UV flux that is minor at $z=0$ but already
dominant at $z\sim 0.3$.

\acknowledgements

Support for proposal 8564 was provided by NASA through a grant from
the Space Telescope Science Institute, which is operated by AURA,
Inc., under NASA contract NAS 5-26555.  We are grateful to
A.V. Sweigart for useful discussions.

\vskip 0.1in
\hskip -0.2in
\parbox{3.5in}{\epsfxsize=3.5in \epsfbox{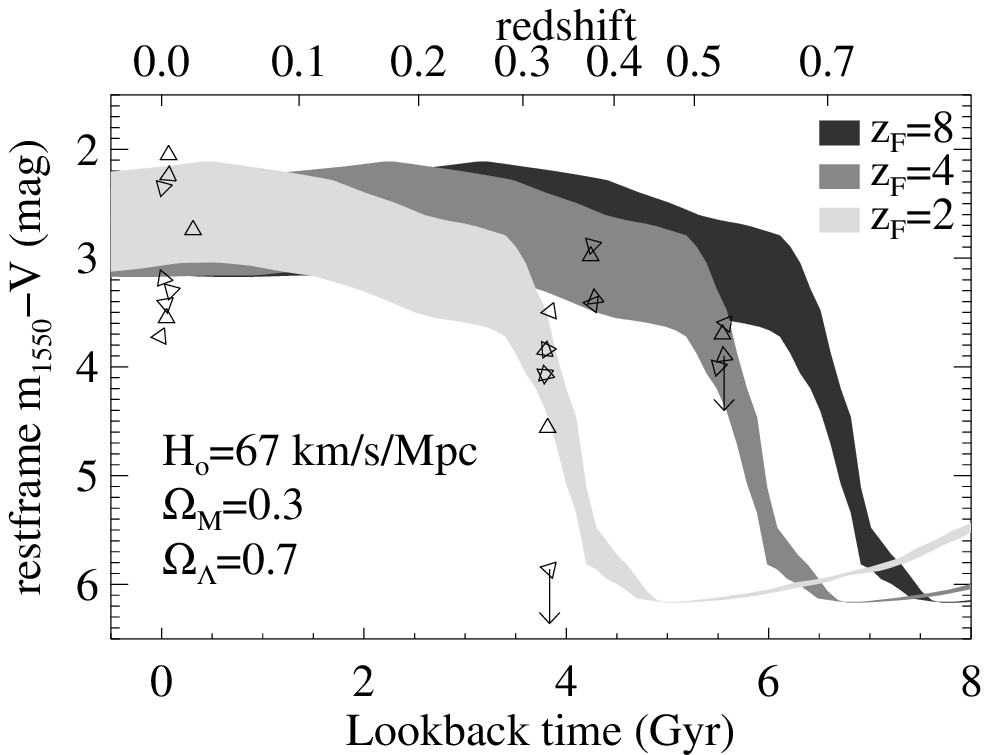}} \\

\centerline{\parbox{3.5in}{\small {\sc Fig.~2--}
Evolution of the
rest-frame $m_{1550}-V$ color according to the models of Tantalo et
al.\ (1996) (shaded), assuming
a reasonable cosmology and three different epochs
of galaxy formation (labeled).  The sudden onset of UV emission is
caused by the appearance of metal-rich hot HB stars, with the spread
in colors bounded by models at $M=3\times 10^{12}$ and $M=10^{12}
M_{\odot}$.  The timing of the onset is much more sensitive to
$z_f$ than to the cosmological parameters.  The triangles represent the
colors measured to date for quiescent E and S0 galaxies.
The statistical uncertainties (Table 1) are small with respect to the
variation from cluster to cluster, except for the 1$\sigma$ limits 
shown (arrows).}}

\end{document}